**5**

# Web Services-Enhanced Agile Modeling and Integrating Business Processes

Fatima-Zahra Belouadha, Hajar Omrana and Ounsa Roudiès
*Siweb (Information System and Web), SIR Laboratory,*
*Ecole Mohammadia d'Ingénieurs, Mohammed V-Agdal University,*
*Morocco*

## 1. Introduction

In a global business context with continuous changes, the enterprises have to improve their productivity and their brand image according to a given strategy. They are expected to face big challenges in order to overcome hard competition and meet the growing demands of clients. Indeed, they should capitalize on the potential of new information and communication technologies to enhance their operational efficiency, to react more quickly and to ensure the flexibility of their business processes. Therefore, they should use e-business methods, mechanisms and techniques. In simple terms, e-business refers to the use of Intranet to network, to enhance and to empower intra-enterprise business processes, and of Internet to serve customers and to collaborate with business partners reusing external processes. Nevertheless, e-business aims cannot be resumed to a simple design of Intranet applications or Web sites to share business data or publish products and business services. It is necessary to note that e-business engages managing business processes. It intends particularly to model, to integrate and to make business processes able to exchange data. Indeed, Business Process Management (BPM) must be done in the most appropriate way to ensure interoperability between both intra and inter enterprises information systems, and to guarantee business process agility and dynamic integration.

The ability to exchange data between heterogeneous systems has been for a long time a prime concern of researchers in this field. Nevertheless, e-business is not limited to a simple exchange of data. Exchanged data are just the product of a set of collaborative business processes identified according to the enterprise business strategy. These processes are implemented by both internal and external systems. Thus, it is necessary to manage both internal and external business processes, implemented on either compatible or heterogeneous systems, while ensuring the data exchange between them. In other terms, it is necessary to ensure the systems' interoperability but it is also important to be able to manage the processes they execute.

BPM aims to discover, design, deploy, execute as well as interact with, operate, analyze and optimize business processes, and essentially to do this at the design level (Smith et al., 2002). Therefore, before deploying end executing a business process, it is necessary to model it as finely as possible at the design-time. Well describing processes is a prerequisite for their strategic alignment. In addition to an expanded specification of business processes, it is also necessary to integrate them in order to achieve a given business goal. However, in an evolving collaborative context, this integration should be dynamic and easy to make the enterprise's information system agile. To achieve this goal in the global economic context, we think that the e-business mechanisms have to provide the means to quickly and easily discover the appropriate business processes through Internet.

In this chapter, we propose a model-driven approach, based on Web services standards, for modeling and integrating agile business processes using Web services. The choice of focusing on Web services technology was not arbitrary. The large and broad adoption of this technology by enterprises will lead most business processes to be performed using Web services. Besides, the added value of Web services and their great interest to business process management are beyond doubt. Web services produce, on the one hand, loosely coupled applicative components. On the other hand, they are the most widely used implementation technology of SOA (Service-Oriented Architecture), which is based on the large experiences of software and distributed component technologies. Being founded on the XML (eXtensible Markup Language) language, the SOAP (Simple Object Access Protocol) protocol and the UDDI (Universal Description Discovery and Integration) repository, this technology can be considered as an appropriate mean to ensure interoperability, data exchange and the publication and discovery of business processes when they can be implemented as Web services. Furthermore, Web services standards developed by the W3C (World Wide Web Consortium) consortium fit well with required standards in e-business context (Belouadha & Roudiès, 2008), and therefore they can be used in case of business processes.

Our approach aims to benefit from Web services characteristics for managing business processes. The purpose is to enable the production of agile business processes through an enhanced modeling and dynamic integration. Therefore, we propose to describe them from a business point of view. We consider that a process is a set of activities that can be mapped to Web services. Thereby, a business process can be implemented by composing a set of published Web services. Besides, agility can be achieved by the ability of updating a given process through reuse of Web services dynamically discovered from Internet. A rich Web services' repository available in the Web as well as the high interoperability of Web services facilitate integration. This chapter is structured in seven sections. Section 2 introduces the concept of business process. Sections 3 and 4 are respectively devoted to the description of modeling languages and integration solutions used to design and integrate business processes. Section 5 presents the proposed approach and illustrates it by a sample scenario. Section 6 explores future research opportunities to build on this work. Finally, Section 7 discusses our contribution by comparing it with related works, and specifies its impact, advantages and also limitations.

## 2. Business process

The term "process" (in Latin "procedere") is defined (in Oxford dictionaries[1]) as a series of actions or steps taken in order to achieve a particular end. Specific and several meanings may be assigned to this concept according to the domain of use: juridical, biological, scientific or technical domain. In the technical domain we are interested in this chapter, a process is often denoted using the term "business process".

---

[1] http://oxforddictionaries.com/definition/process

A business process denotes a set of activities to be performed by given actors through specific tools and methods according to a defined procedure. Many definitions have been conferred on this term. For example, Hammer et al. define an operational process as a sequence of activities which, from one or several inputs, produces a result (output) that represents an added value for a client (Hammer et al., 1993). Lorino considers a process as a set of activities connected together, using information flows and combined to provide an important and well-defined material or immaterial product (Lorino, 1995). For Tarondeau, a business process is a set of activities which are organized into network, in a sequential or parallel manner, and which combine and implement multiple resources, capabilities and competences to produce a result or output of value for an external client (Tarondeau, 1998). Brandenburg and Wojtyna consider it as a chain of activities or activity sets, which is supplied by inputs, has resources and adds value relative to the goal in order to create outputs (Brandenburg and Wojtyna, 2003). Besides, Morley et al. use five notions: goal, activity, role, resource and event to define a business process (Morley et al., 2011). They identify it as a set of activities undertaken in a determinate objective, assigned in total to an actor or in part to several actors corresponding to different roles, performed using resources, and may be conditioned by events of internal or external origin. They also introduce the notion of process structure that corresponds to the arrangement of process activities. Finally, the ISO 9000 standard defines a business process as a set of correlated or interacting activities which transforms inputs into outputs. All of these definitions commonly agree on the fact that a process is a set of activities which, from a set of inputs, produces a set of outputs that allow achieving a given goal. Moreover, we note that activities are performed by human actors or machines and can require resources. They are sometimes conditioned by events. Moreover, they can also interact and they run in a determinate order according to expected events. To specify business processes, it is therefore necessary to use concepts such as activity, input, output, goal, role, resource and event.

In the literature, an activity refers to tasks. Lorino identifies it as a set of elementary tasks, homogeneous from the point of view of their cost and performance behaviors, which produce an output from a set of inputs (Lorino, 1991). An input is an object on what the activity must operate to produce an output. The output is simply a result produced by the activity. The goal of a given process is what this process aims to achieve or to accomplish in order to meet the enterprise strategy. The role refers to a prescribed or expected behavior associated with a particular status in the enterprise. It corresponds to a responsibility assigned to an internal or external actor which can be a person, an organizational entity (which includes a group of persons) or a system. We can consider that a role corresponds to one or more activities that are performed by the same actor. However, an actor may play several roles. It is therefore appropriate to specify a business process by eliciting the roles rather than the actors. Besides, the concept of resource refers to a mean required to perform an activity. It can be for example a database, a software or a hardware tool. Last but not least, the event constitutes a fact that occurs and which can trigger an activity. It can, for example, correspond to a time deadline or an occurring result (output).

Finally, we note that the activity is the core of a business process. We can therefore apprehend this one as a decomposable macro-activity and consider the task as the smallest unit of decomposition. The process approach is, in fact, a systemic approach that distinguishes several levels of analysis. These ones can be, in general, summarized into four levels (Brandenburg & Wojtyna, 2003): the macro-processes, the elementary processes, the sub-processes that we choose to call micro-processes in this chapter trying to avoid any confusion with the term sub-processes used in the BPMN (Business Process Management Notation) standard (Object Management Group [OMG], 2011), and finally the activities. A macro-process provides added value and meets a strategic goal of the company. A multi-activity enterprise can use several macro-processes. Each macro-process is, in fact, decomposed into elementary processes, and each elementary process is simply decomposable into activities or even micro-processes when its activities are performed by different entities. In this last case, each group of activities constitutes a micro-process. We can say that a micro-process is simply a part of a business process composed of activities, performed by a same entity and hence corresponding to a same role. After this analyze of the process concept, let's consider in the following section its representation in the literature.

## 3. Languages and formalisms for modeling business processes

Modeling business processes has, for long time, been attracting much research interest. Many languages, formalisms or even frameworks have been developed over time to meet that goal. We cite, as examples, Petri nets, the CIMOSA (Computer Integrated Manufacturing Open System Architecture) framework, the RAD (Role Activity Diagrams) notation, the EPC (Event Driven Process Chain) notation, the IDEF (Integration Definition Language) methods, the OSSAD (Office Support Systems Analysis and Design) method, the Merise method, the UML (Unified Modeling Language) language and the BPMN notation. In this section, we study these formalisms in order to show their fundamentals as well as their limitations. However, we cannot describe all of them here. We focus on three standard languages: IDEF, UML and BPMN. We have chosen these three languages because we think that they trace the evolution in modeling business processes over the years. IDEF has been the first brick proposed in this area. UML has accompanied the object-oriented paradigm and has been widely adopted by designers. Finally, the BPMN notation is a recent standard that is known to be the most adapted standard to model business processes.

### 3.1 IDEF

In 1970's, U.S. Air force has launched a project for integrated computer-aided manufacturing. This project has led to the IDEF family of methods. Today, this family includes fifteen methods recommended to describe and specify business processes. The first two methods, IDEF0 and IDEF1, known as SADT (Structured Analysis and Design Technical) (Lissandre, 1990), have been developed to respectively represent the functions and information. IDEF1 focuses on describing company data models, whereas IDEF0 has been devoted to the description of the processes.

IDEF0 had great success and became an IEEE standard in 1998. It considers that a complex system is iteratively decomposable into simpler subsystems that interact to achieve a goal. This decomposition consists in splitting the system into a set of functions, starting from a global function to dissect into sets of sub-functions, until reaching elementary functions. IDEF0 provides a structured system representation with hierarchical levels. It adopts a graphical notation and a simple syntax in order to describe and organize the system in a tree. This method represents in fact, a business process as a tree whose nodes constitute functions described using diagrams. Each diagram includes one or more functions which constitute nodes of the business process tree and belong to a given level of decomposition.

To describe a function, IDEF0 considers it as an activity or a sub-process that transforms ingoing objects into outgoing ones, through mechanisms that correspond to material, software or human means (resources, tools, actors, etc.). It also assumes that its behavior can be influenced or triggered by a set of conditions that form its control directives. Indeed, IDEF0 describes each function as a box, expresses its goal simply using a text and represents its interface using arrows indicating its inputs, outputs, mechanisms and controls. A business process model would thereby be modeled as shown in Figure 1. A node number can be mentioned under a box in the IDEF0 diagram to indicate that the corresponding function is a composite activity which is decomposed and described in another diagram denoted by the same node number. In Figure 2, *Node 0* is described by a graph with two activities and a sub-process (*Node 2*).

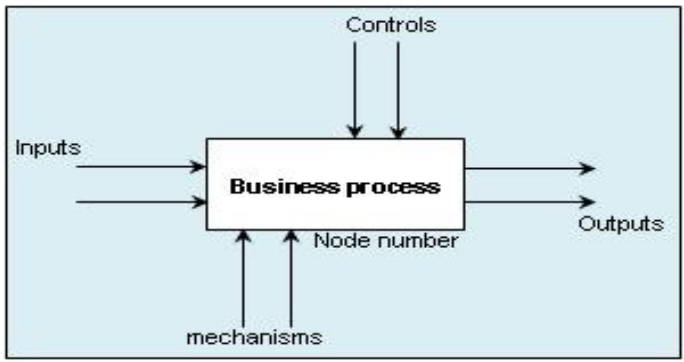

Fig. 1. IDEF0 Business process model.

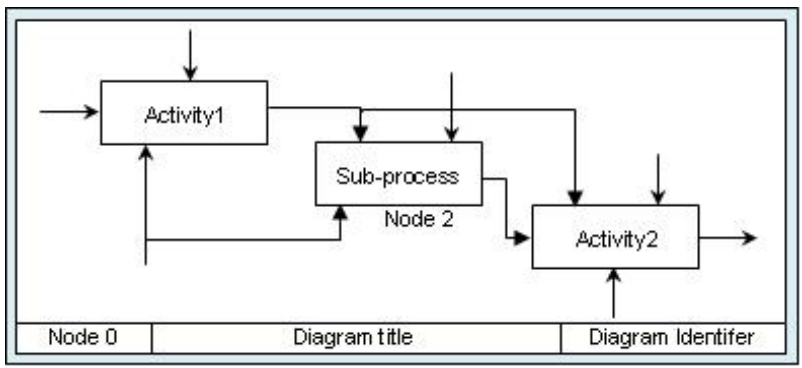

Fig. 2. Sample model of a diagram decomposing a node in IDEF0.

Although the IDEF method had remarkable success in industrial and military domains, it has some limitations. On the one hand, IDEF0 does not provide a flexible way to describe the process activities and requires describing each box (activity) by at least an output and a control. It also leads to an ambiguous description of concepts: since the actors, roles and resources are all represented by ingoing or outgoing arrows, it is not always easy to distinguish, for example, an actor from a role, or even an actor from a resource. On the other hand, IDEF3 does not allow building a well-developed representation of the process behavior, and informally describes any additional information concerning a process by simple comments. Several years later, UML has focused on non ambiguous description of systems.

**3.2 UML**

UML is an object-oriented modeling language (J. Gabay & D. Gabay, 2008; Charroux et al., 2010). It was created by merging the methods OMT (Object Modeling Technique), OOD (Object Oriented Design) and OOSE (Object Oriented Software Engineering), and became an OMG (Object Management Group[2]) standard in 1997. UML Version 2 proposes thirteen diagrams to describe a system. To formalize a process, it is useful to use, among them, the use case diagram, the transition-state diagram, the activity diagram, the sequence diagram, the communication diagram, the interaction overview diagram and the time diagram shown in Figure 3.

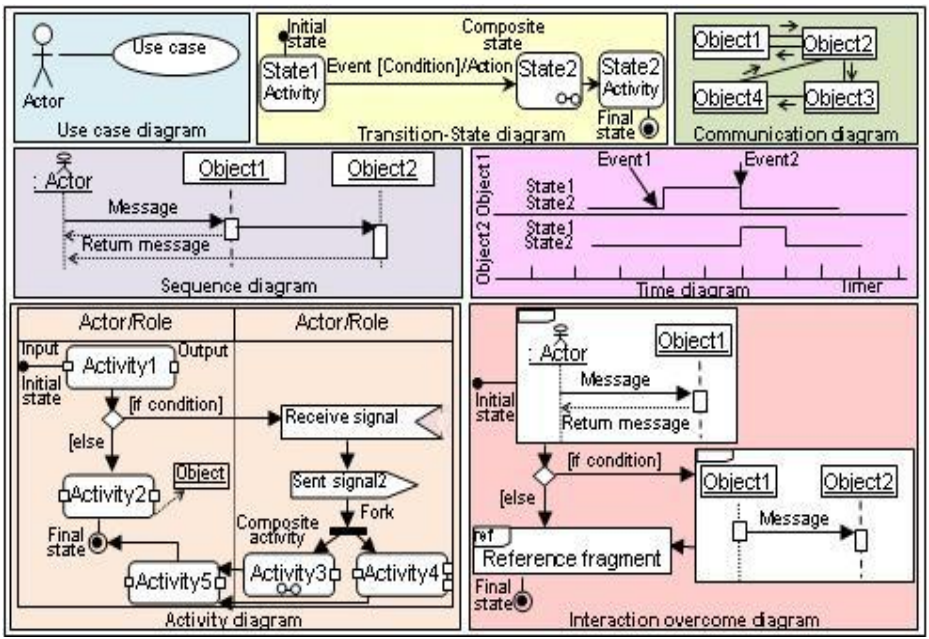

Fig. 3. View of UML behavioural diagrams used to describe business processes.

The use case diagram describes a system from the point of view of the user. The state-transition diagram represents the lifecycle of an object and its main states. Just as the transition-state diagram, the UML 2 activity diagram also illustrates the behavior of the system by introducing new elements to provide a rich description of activities and their sequence. It specifies the activities using input/output pins, describes data flows (objects), and uses specific nodes to describe control flows such as decision, fork and synchronization. It also introduces specific communication actions to represent events such as a signal or a time flow. In addition, it can be organized into partitions, each one including a group of activities that often correspond to organizational units or actors. Finally, the activity diagram is the core of process description, in which micro-processes can be elicited using partitions. Besides, the sequence diagram is a temporal description of a collaboration scenario between objects. The communication diagram, as for it, only indicates interactions between objects. The interaction overview diagram merges the activity and sequence diagrams to provide more detail. Finally, the time diagram allows to better visualize the state changes and the interactions between objects in respect of time constraints.

---

[2] www.omg.org

All of the diagrams mentioned above can be used to model macro-processes in UML. The use case diagram serves to show the use of elementary processes by different actors. These elementary processes can be further specialized in micro-processes to provide more details. As for the activity diagram, it is used to formalize the activities and tasks of each elementary process. The aim is to describe the control flow between all of the micro-processes constituting an elementary process, since their activities are interdependent. This description could be performed considering as many partitions as there are actors, or even micro-processes. Besides, the communication diagram can be used to describe interactions (exchanges) between micro-processes, and the sequence diagram would be useful to precisely describe their chronology. In addition, the interaction overview diagram improves the visibility of each elementary process conduct by merging the activity and sequence diagrams. The object state transitions, as for it, can be described using the state-transition diagram, but also the time diagram when these states change over time depending on their interactions with other objects.

Finally, we note that modeling a business process is spread over seven UML diagrams, each one provides an accurate modeling focusing on a given aspect. Therefore, the activity diagram, which is fundamental for specifying the process behavior, does not cover all of the elements describing a process such as the events or resources. It denotes a resource as a simple comment, and does not show the events triggering activities, except for communication actions which can be considered as signal type events. The events are, rather, described at the level of the state-transition diagrams in which the activities are encapsulated in objects states. Thus, the activity diagram cannot be directly mapped to an executable process. In addition, details on the processes' inputs and outputs can only be given through an eighth diagram: class diagram. However, this one only provides information about the data types of these parameters. After these two generic process notations, let's consider in the following sub-section BPMN standard which is devoted to business processes design.

### 3.3 BPMN

BPMN is a flowchart representation of business processes. It aims to provide, on the one hand, a fully graphical notation, easy to use and usable at both business and technical levels, and on the other hand, a mechanism to generate executable processes directly without using another language. It was developed at the initiative of the BPMI (Business Process Management Initiative) organization in 2004 and adopted as OMG standard in 2006. Version 2 of the BPMN standard (OMG, 2011) uses five main concepts for this purpose: the swimlanes, the flow objects, the connectors (connecting objects), the data and the artifacts. As illustrated in Figure 4, all of these concepts are used together into a specific diagram to describe in particular the dynamic behavior of a business process.

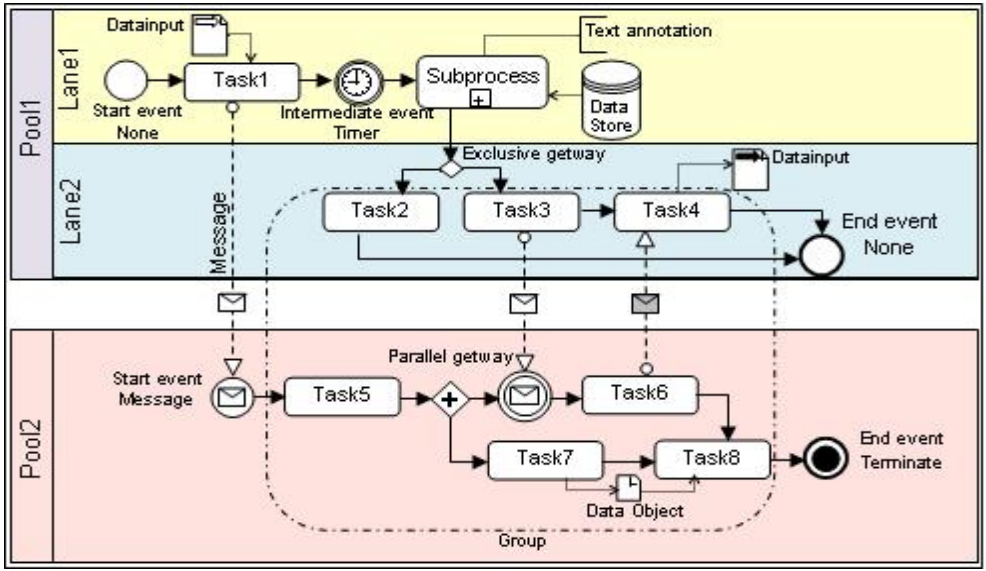

Fig. 4. Sample model of BPMN diagram used to describe business processes.

The swimlanes (pools and lanes) are used to create activity partitions in order to organize processes. Each pool includes activities performed by a participant in an interactive process. The pools are activities' containers, whereas the lanes are a way to organize these activities within the pool. Besides, the flow objects constitute graph nodes that represent the activities (tasks or sub-processes), events and gateways. The tasks are atomic activities and the sub-processes are composite activities. The events are something that happens and affects the process execution. They often have a cause and a result, and can start, interrupt or terminate a flow. The gateways are elements representing the control of activities flows. They are divided into exclusive, inclusive, parallel, event-based, and complex gateways, to represent different conditional branching such as "Fork", "Join" and "Merge." The connectors are information flows that represent sequence flows, message flows, associations or data associations. The sequence flows link the activities within a lane to explicit the order in which they will be executed. They connect together activities, events and gateways, and can be conditional flows. The message flows represent the messages exchanged between the pools. The data associations are used to model the data flow between data objects and activities or events. They can, for example, associate inputs or outputs with an activity. The associations express relationships between information or artifacts and flow objects. For example, an annotation (user-defined text) can be associated with a graphical element. The artifacts (groups or text annotations) provide additional information about the process. The groups are used to highlight sets of activities that may belong to different lanes and/or pools. The text annotations are associated with the BPMN diagram elements to describe them. Besides, the data refer to data objects, data inputs, data outputs, data stores or messages. The data inputs and data outputs are respectively input to/output of the entire process as whole. They can be associated with its activities using data associations. As for the data objects, they represent either objects used by the activities, or objects that the activities produce (business documents, emails, etc.). The data stores are places where the activities can retrieve or update stored information, e.g. databases. Finally, the messages describe the contents of a communication between peers of participants.

### 3.4 Conclusion

BPMN is a standard inspired from a myriad of methodologies and notations such as UML, IDEF, ebXML BPSS (Electronic Business using eXtensible Markup Language Business Process Specification Schema), RosettaNet, and EPC. Nowadays, it is considered as the most adapted notation to business process specification. This standard proposes a set of concepts that can be described in a concise, clear, but also compact (not fragmented) manner. The resulting descriptions are understandable by all business users, and produce interoperable process management systems. The main advantage of BPMN is that it provides processes representations that we can directly implement and execute without having to perform translations.

In this section, we analyzed how to represent business processes and we conclude with the leading position of BPMN. In the next section, we will consider the ways to integrate business processes.

## 4. Business process integration solutions

The integration consists in adopting a technique to ensure the coherence of an information system constituted of different components, create collaboration pathways, reuse functions, share data, and ensure the agility of the enterprise information system. Various solutions have been experimented to achieve the integration of applications. They can be divided into seven classes, each one uses a specific technique: the conversion, standardization, middleware, EAI (Enterprise Application Integration), ERP (Enterprise Resource Planning systems), Workflow and BPEL (Business Process Execution Language) based solutions.

### 4.1 Conversion

The conversion consists in using peer to peer converters in order to convert the models and data exchange formats between applications. This type of integration is chronologically the first integration technique that was specifically used for data exchange. Using a converter allows converting the sender's data format into the receiver format. However, it requires using as many converters as there are used formats. Thus, exchanging data with an application which uses a new format requires conceiving a new converter. In addition, the updating of the data format adopted by a given company application involves the updating of all converters concerning this application. This technique leads to a complex interfacing. Furthermore, it assumes that partners are known in advance, and thereby it is adapted to A2A (Application to Application) integration and not B2B (Business to Business) one.

### 4.2 Standardization

The data integration solutions based on standardization have been developed to avoid the problems of the conveters-based integration. They specifically consist in unifying representation and data exchange models. This class includes, as examples, the EDI (Electronic Data Interchange) standards, and recently the XML-based standards, RosettaNet and ebXML.

The EDI standards like X12 (American intersectoral standard) and EDIFACT (Electronic Data Interchange For Administration, Commerce and Transport) have opted for standardization as an efficient mechanism for intersectoral and international electronic data exchange. These standards have attempted to standardize, according to the domain context, the way, the means and the format used to exchange data. Their aim was to enable partners, belonging to the same sector, to communicate. However, they were designed to support business transactions between a limited set of known trading partners (Albrecht et al., 2003). They also do not provide a standardized representation of information supporting different areas and do not allow indexing for discovery purposes (Truman, 1998).

The standards proposed in 1998 by the RosettaNet consortium, help trading partners collaborate through commercial transactions based on a formalized data exchange. They are developed to be used in a layered infrastructure that partners must necessarily adopt to be able to collaborate. These standards are used to define business process templates, partner interface processes, architecture of exchange and transaction dictionaries. However, they are adapted to e-commerce transactions and require adapting used systems according to the proposed infrastructure.

The ebXML standard was proposed in 1999 by the UN/CEFACT (United Nations Centre for Trade Facilitation and Electronic Business) and OASIS (Organization for the Advancement of Structured Information Standards). As recommended by ISO 15000, it uses five layers of data specification (Chauvet, 2002). These layers are used to describe business process interfaces, define collaboration protocol profiles and agreement, specify reusable data structures, define message transport and routing, and indicate mechanisms required to use repositories and registers. The ebXML BPSS specification uses activity diagram concepts to formalize a business process. It describes it as a set of choreographed business transaction activities which exchange XML business documents to perform, in particular, binary collaborations. The ebXML standard is adapted to business interchanges between commercial partners. It is also a non flexible solution which assumes that multi-party collaborations are very restricted. Furthermore, it is intended for exchanging documents which match with predefined data patterns, and imposes agreed interchanges using only predefined interaction patterns.

**4.3 Middlewares**

The middlewares are used as intermediate software components that make applications communicate and abstract the heterogeneity of their platforms. They mainly ensure messages transport and routing, but also other services such as data transformation and transaction management. Middlewares, such as JDBC (Java DataBase Connectivity) and ODBC (Open DataBase Connectivity), ensure data access regardless of the databases used. Those such as API RMI (Java Remote Method Invocation) and MOM (Message-Oriented Middleware) are based on the RPC (Remote Procedure Call) techniques or message exchange to make communicate remote systems. Besides, the component-based middlewares are among the most important ones. They constitute a solution to integrate heterogeneous and distributed applications. Being based on remote methods or procedures invocation, they make communicate, according to specific communication models, components distributed across network. Examples of this type of middlewares are CORBA (Common Object Request Broker Architecture) of OMG, COM/DCOM (Component Object

Model/Distributed Component Object Model) of Microsoft, J2EE (Java 2 Enterprise Edition) of Sun and DotNet of Microsoft.

Finally, the component-based middlewares have, in general, the advantage of ensuring portability, security, scalability, load balancing, and reusability. However, they have limitations that can be summarized in the complexity of implementation, the cost of platforms used, the tight software coupling that makes the produced applications non flexible, the dependence on the development environment due to the use of specific communication buses, and therefore the complex integration of heterogeneous systems. Furthermore, some of these middlewares are proprietary and non standardized softwares.

### 4.4 EAI

EAI provides formatting, technical gateways and process management services (Octo Technology, 1999). They allow, in general, conveying messages, accessing applications through interfaces provided by connectors, transforming and conveying data in a format adopted by the target application using a pivot format, ensuring messages routing and also orchestrating applications in the case of some EAI. BEA Weblogic Integration, Microsoft Biztalk and IBM Websphere Business Integration are among the EAI technologies used in the market. These technologies have the advantage of integrating heterogeneous applications using connectors, and reducing the connections between these applications by centralizing them at the EAI level. However, the connectors are proprietary and difficult to maintain. In addition, implementing EAI is a complex operation, and the interoperability they provide depends on the EAI which constitutes a central intermediate.

Besides, the ESB (Enterprise Service Bus) technology, proposed by the Gartner Group in 2003, constitutes a new generation of EAI. It uses diverse standards such as XML, JMS (Java Message Service), JCA (Java Connector Architecture), JMX (Java Management Extensions) and also Web services standards. An ESB provides a set of services (Mullenders, 2009): dynamic services discovery, their automatic orchestration, strong distribution services on the network or Internet, messages communication, XML data transformations, content-based intelligent routing, and in some case, business Activity Monitoring, and business process management. Several ESB exist in the market. They are proprietary as well as open source, such as: IBM WebSphere ESB, Microsoft BizTalk Server, Oracle Enterprise Service Bus (BEA Logic), Apache ServiceMix and Open ESB. The ESBs are based on bus architecture, exploit Web services and use an integration engine which is distributed in the service adapters. It follows that they provide highly distributed integration and lead to loosely coupled applications. However, the administration of these tools may be complex depending upon the integrated systems. Besides, they are more costly, vendors depending, and require more work upfront. They also assume the use of unified messages models. Finally, the ESB may become a single point of failure since all application communications are performed across it. These indirect communications may lead in increased latency and decreased performance.

### 4.5 ERPs

ERPs are computer systems which use advanced communication and information technologies to process enterprise management information. They constitute, indeed, a solution that performs all of the crucial enterprise business processes such as the commercial management of the supply chain, from product production to its sale. They consist in a set of independent applicative modules and use a workflow engine to propagate real-time data between these modules. SAP, Oracle/Peoplesoft, SAGE ADONIX, and Microsoft are examples of proprietary ERPs. As for Compiere/Adempiere, Ofbiz, Openbravo and Open ERP (previously denominated Tiny ERP), they are among Open ERPs. All of these systems are considered as the most popular ERPs in the market.

Besides, ERPs have, in general, the advantage of overcoming the difficulty of information access due to the incompatibility of platforms, databases, data formats and semantics used by different business applications. They are among the solutions adapted to A2A process integration. However, they require major investments (high costs of license, maintenance, technical assistance, training and material infrastructure), a dependence of the supplier in case of technical problems and a complex implementation. Implementing an ERP project requires, in fact, to conduct an analysis and preliminary study in the long term to perform the ERP customization, update it if the version has been changed, create appropriate interfaces to integrate the ERP with existing applications, and make organizational changes and adaptations according to the business processes defined by the ERP.

### 4.6 Workflow

Workflow engines aim to automate business processes by automating the coordination of the tasks and the circuits routing documents as well as information between a set of actors. We cite, among them, Windows Workflow Foundation, WORKEY and Activiti . They allow executing one or more workflow definitions described using workflow languages. These languages are used to represent a process by an activity flow specifying the exchanged documents and information as well as the actions of each actor.

WPDL (Workflow Process Definition Language) is, for example, a workflow definition language that was defined by the Workflow Management Coalition (WfMC) in 1994. It is based on a meta-model (objects model) and allows describing the basic elements of a workflow management system with a focus on the interfaces to external systems. Its use can enable the exchange of models and documentation between heterogeneous worlflow systems of different vendors (Workflow Management Coalition [WfMC], 1998). For the purpose of using an internationally defined standard encoding language, WPDL uses today XML to exchange processes, and is denominated XPDL (XML Process Definition Language). However, this language is not a formal language and does not provide flow semantics, and thereby, its implementation and use impose problems. In addition, although the workflow languages enable to easily define a workflow between human actors, the integration of the workflow to an application or a system still remains a complex task.

### 4.7 BPEL

The BPEL language, originally called BPEL4WS (Business Process Execution Language for Web Services) is a recent language appeared in 2003. It was inspired from the WSFL (Web Services Flow Language) and XLANG (XML Business Process Language) standards. This language is intended to execute synchronous and asynchronous business processes. Besides, it is fundamentally based on the use of XML standard to describe an executable process with all the necessary technical specifications. It is also worth noting that this language allows the rapid and easy deployment of executable processes on a business process engine.

Version 2 of BPEL constitutes an OASIS standard appeared in 2007 (Organization for the Advancement of Structured Information Standards [OASIS], 2007). This standard allows interacting with Web services to exchange XML data, and managing exceptions and compensation in the case of rollback transactions. It describes the process activities using flow control structures (sequence, iteration, parallelism, etc.), defines the data manipulated in containers and organizes Web services, when used, as partners. In the BPEL context, each activity consists in a service invocation, a message reception or transmission, or an exception (error message) reception. As for the exception and compensation routines, they are respectively described using specific elements, namely, *faultHandlers* and *compensationHandler*.

### 4.8 Conclusion

One main feature of BPEL is that it constitutes, in fact, the standard language for Web services composition. Thereby, it is a recent and recommended standard for orchestrating transactional and distributed business processes. It is, moreover, adapted for the specification of executable as well as abstract business processes, in particular, those apprehended as Web services. That's why we adopt this language in combination with BPMN standard in order to model and integrate business processes. The main lines of our approach for business processes are introduced in the next section.

## 5. Our approach for modeling and integrating business processes

In this section, we propose a Web service and model-oriented approach for modeling and integrating business processes. The aim is to produce flexible and scalable systems. These systems would be able to use agile business processes that can be dynamically integrated and subsequently, become adaptable to changes. To achieve this goal and realize efficient information systems, we think that enterprise must adopt a development cycle fundamentally based on modeling and integrating business processes. It could begin with designing its abstract business processes according to its needs and regardless of its competences, then implement or simply discover Web services that can be used in order to perform those processes if it decides to collaborate with business partners. Finally, it is this discovery aspect in which we are interested, since it leads to a dynamic integration producing changeable business processes. However, dynamic services discovery, in this case, can only be performed if the business processes design covers their functional, non-functional and semantic aspects. Eliciting what a process exactly do, its non-functional properties (e.g., security and quality of service), and the semantics of its inputs/outputs as well as its functionality, is crucial to manage processes, and in particular to discover and integrate them. Thereby, three main phases define our global approach: business process modeling, dynamic Web services' discovery and deployment.

The initial modeling phase consists, first of all, in describing all abstract business activities making up the global process, using UML and BPMN diagrams and taking advantage of the concepts used in Web services standards. It must obviously cover the business process's functional, non-functional and semantic aspects. This phase delivers generated BPEL files corresponding to the abstract business processes. The second phase of dynamic Web services discovery consists in searching, among Web services published in repositories, those that are appropriate to automate the abstract business processes. As for the deployment phase it simply consists in automatically replacing each invocation of an abstract activity in BPEL files generated in advance, by a concrete invocation of the corresponding Web service. The objective is to make business processes executable by adding the relevant technical information (binding information in the Web services context) to their BPEL files. In the following section, we explain the details of our approach concerning the two main phases: business process modeling and Web services dynamic discovery. We first present the modeling principles that will be illustrated using a sample scenario. Then, we discuss the proposed discovery mechanisms.

### 5.1 Business processes modeling

The BPMN standard adopts an activity-oriented representation for specifying business processes. This type of representation is adapted for describing the behavior of processes. However, it does not describe their functional, non-functional and semantic aspects. In fact, even if the BPMN standard introduces elements such as the goal and inputs/outputs to specify a business process, it does not describe it in an expanded and disambiguated manner. In this subsection, we propose an approach that exploits the concepts used in the W3C standards recommended to describe Web services, in order to model business processes. This approach does not ignore the importance of the activity-oriented representation proposed by BPMN. It, indeed, adopts it to model the behavioral aspect of business processes, but enriches and completes it by functional, non-functional and semantic representations using UML diagrams. Our business processes metamodel is illustrated in Figure 5. The metaclasses shown in yellow, blue, and pink respectively refer to concepts used to achieve the behavioral, functional and non functional description while considering the semantic of business processes. By analyzing all of the definitions and concepts previously presented, a business process may, indeed, either represent a micro-process, an elementary process or a macro-process. A macro-process is an aggregation of elementary processes, each one can aggregate one or more micro-processes. Each micro-process is an aggregation of activities and the activity is specialized into a task or a subprocess. According to the BPMN standard, we denote a task as an atomic activity and a sub-process as a composite activity. We have, therefore, used the composite pattern to specify the metaclass *Activity*.

As we previously noted, we are considering to specify business processes as Web services. A main question remains: how to introduce this technology in the business processes specification practices? To answer this question, we think that this can be performed through the business processes' activities. We consider that each activity part of a business process refers to an atomic or composite Web service, and each task corresponds to an operation. This mapping is justified by the fact that an activity, just as a Web service, includes a set of operational tasks (operations) that cooperate together to provide a specific function and produce added value. Thereby, the atomic activity can be performed by simply invoking an operation of an atomic Web service. Whereas, the sub-process is a composite activity that can be performed by an atomic or composite Web service, and thus requires orchestration of a set of operations of the same Web service or coordination of several Web services.

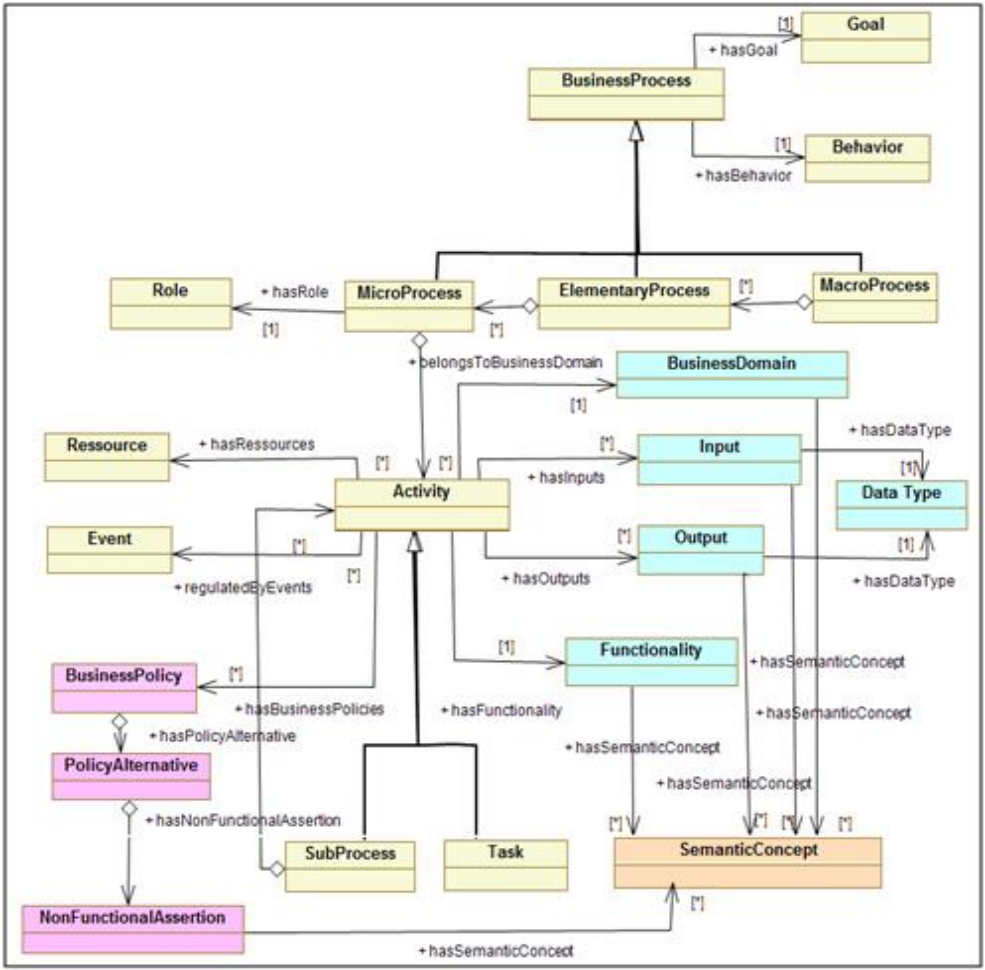

Fig. 5. Business process metamodel.

Besides, we recall that a business process can be described by a goal, activities, roles, resources, inputs/outputs and eventually a set of events. However, a global process is represented by its goal and eventually its decomposition into processes (Morley et al., 2011). The macro-processes and elementary processes are elements of great granularity from the point of view of analysis. They are, in reality, global processes that do not need to be described in detail. The details of their specification may in particular be given at the level of their fine-grained components. Thereby, it would be convenient to associate the metaclasses *goal* and *behavior* with the metaclass *business process*, in order to describe indifferently a macro-process, an elementary process or a micro-process by their goals as well as their behavior. Besides, the metaclass *role* should rather be associated with the micro-process which by definition includes a set of activities assigned to the same role. The micro-process specification also requires identifying its inputs/outputs, resources and events to which it reacts in order to understand its running. However, for a detailed description, these elements, represented by the metaclasses *Input*, *Output*, *Resource* and *Event* must be specified at the activity level to describe their implementation. Each input or output should also be described by its data type and each activity should be described by its business domain and its functionality. All of these elements will serve to specify an abstract business process which is not yet associated with one or more concrete Web services to become executable.

The W3C standard WSDL (Chinnici et al., 2007) specifies a Web service as an interface that belongs to a business category, and aggregates operations providing a given functionality and having input and output parameters. Our way to specify the business processes allows describing their functional aspect, while fitting harmoniously with this standard. It also introduces the concepts role, resource, event and behavior (orchestration in Web service context) to describe their dynamic aspect. This aspect may, indeed, be represented by a BPMN diagram. This one uses different workflow patterns to describe the arrangement of a business process, while organizing activities by role and highlighting events and resources. However, this is not enough to describe the semantics and non-functional properties of business processes.

To specify the business processes non-functional aspect, we reuse the concepts introduced in WS-Policy standard (Vedamuthu et al., 2007) recommended to describe Web services' policies. According to this standard, we complete the description of an activity by a business policy. We model this activity as an aggregation of alternative policies, each one constitutes an aggregation of assertions. A business policy refers, in fact, to non-functional properties that must be satisfied during the execution of an activity. It is expressed by mentioning sets of alternative non-functional properties. Each of these sets constitutes an alternative policy and each non-functional property is an assertion in the sense of WS-Policy. The Web service that will perform an activity must therefore have, as non-functional properties, all the assertions belonging to one of its alternative policies. Besides, to produce semantic business processes, we take our inspiration from the SAWSDL standard (Farrell et al., 2007). By analogy to the *ModelReference* element, we use the metaclass *SemanticConcept*. We associate this metaclass to the metaclasses *BusinessDomain*, *Functionality*, *Input* and *Output* which describe the business process functional aspect, and also with the metaclass *NonFunctionalAssertion* that specifies its non-functional aspect. The metaclass semanticConcept will allow describing all of these elements using ontological concepts that are semantically equivalent to them, and thereby will serve to dynamically discover Web services able to perform the concerned business processes.

### 5.2 Sample scenario

In this subsection, we present a scenario of selling silver by an enterprise supplier S to illustrate the use of our approach. We assume that when the enterprise S receives a new order from an enterprise client C, it conducts the preparation of the ordered quantity of the silver. Once the supplies are ready, it needs to identify the international current price of silver and convert it into euro before billing the order. Once the invoice is ready, the company C receives the invoice and supplies and must make the payment on delivery. Besides, we assume that the identification of the real-time price of silver and its conversion into euro according to the current exchange rate cannot be performed internally. Thereby, the enterprise needs to use external services provided by an eventual partner. It should indeed be able to discover Web services that provide these financial services and dynamically integrate them to its system.

As defined in this scenario, we can classify the business process activities performed by the company S into three categories: Internal, external and abstract activities. Internal activities are implemented by the information system of the company S, and consist of three activities: *ObtainSilverOrder*, *DeliverSilver* and *ReceivePayement*. External activities are performed outside the company S. They are implemented by the information system of the client C. These activities, denominated *Place SilverOrder*, *ReceiveSilver* and *Pay*, are statically integrated in the process of company S. Finally, abstract activities have to be dynamically integrated into the related business processes by discovering the appropriate Web services able to automate them. They consist of two financial services: *GetRealTimeSilverPrice* and *CurrenciesExchange*. To describe this scenario, we use the BPMN notation as shown in Figure 6.

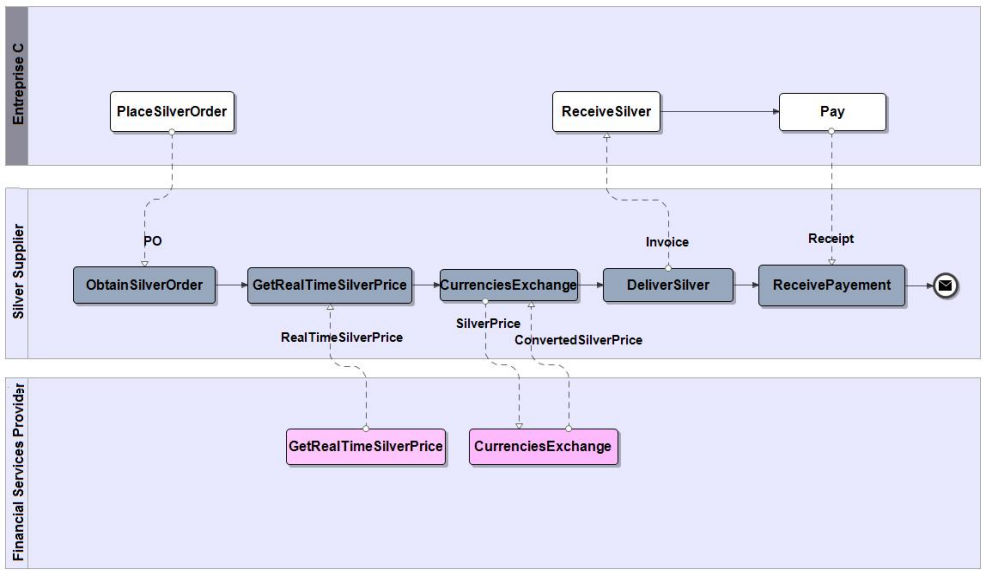

Fig. 6. Process of purchasing silver's metal.

The illustrated BPMN diagram (Fig. 6) is used to model the behavior of the business processes required by the enterprise S. It identifies three pools representing the concerned company, its customer and its external partner. The pool Silver Supplier represents the company S which constitutes the silver's supplier. The pool Enterprise C refers to the business customer of this one. Finally, the pool Financial Services Provider represents a partner that provides financial services and which has to be identified by the company S through discovery process. Each of these pools contains the activities performed by one of the three entities: the company S, the customer C and the abstract partner.

In order to dynamically discover Web services able to perform the abstract activities *GetRealTimeSilverPrice* and *CurrenciesExchange*, the company S must complete its BPMN diagram by two UML diagrams. These ones cover the description of the functional, non-functional and semantic aspects of the concerned abstract activities, in accordance with the metamodel previously exposed. In the remainder of this subsection, we present the case of the activity *GetRealTimeSilverPrice*. Modeling the second abstract activity *CurrenciesExchange* can be done similarly. Figure 7 illustrates an object diagram that specifies the functional properties of the *GetRealTimeSilverPrice* activity, and their corresponding semantics. The concerned activity is, indeed, described by its business domain, functionality, output and ontological concepts which constitute semantic annotations of these properties.

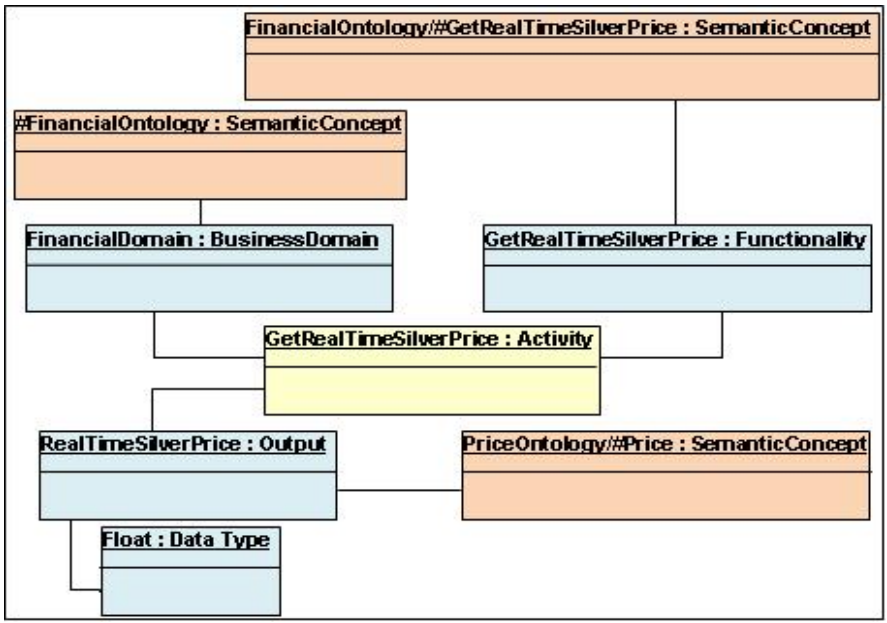

Fig. 7. Object diagram describing functional aspect of *GetRealTimeSilverPrice* activity.

Finally, we consider that to ensure a secure message exchange between the Web service performing the abstract activity *GetRealTimeSilverPrice* and the other activities of the company S, a security policy is required. This policy consists in using cryptography or digital signature to secure the exchange between all involved partners. Modeling this non-functional aspect is thereby essential to discover the appropriate Web service that meets this non-functional need. Figure 8 illustrates an object diagram that describes the involved non-functional constraints and their corresponding semantics. *GetRealTimeSilverPrice* activity is associated with a security policy that aggregates two alternatives policies: *Alternative 1* and *Alternative 2*. The first alternative policy requires the adoption of cryptography mechanism to secure the exchange using the assertion *EncryptedParts*. As for the second alternative policy, it requires the adoption of digital signature mechanism using the assertion *SignedParts*. Both *EncryptedParts* and *SignedParts* assertions are predefined in the WS-Security standard. In our example, they are annotated by ontological concepts to link them to their semantics.

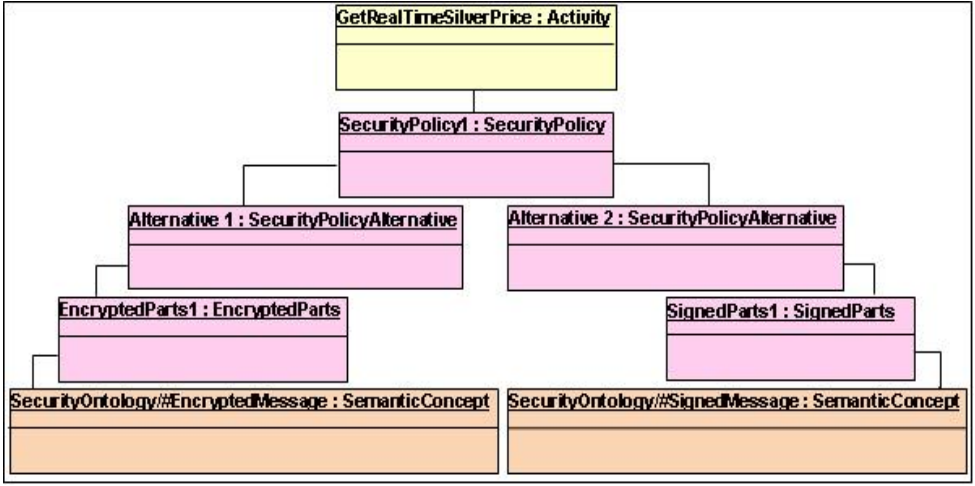

Fig. 8. Object diagram describing non-functional aspect of *GetRealTimeSilverPrice* activity.

In this section, we considered a sample scenario in order to illustrate how our approach designs business processes. In the next one, we address the Web services discovery issue.

**5.3 Web services dynamic discovery**

The second phase of the development cycle we propose consists in Web services dynamic discovery. This phase involves three basic steps: extraction, matching and selection. It consists, first, in extracting, from published WSDL files, the functional, non-functional and semantic properties of the corresponding Web services. To determine the set of Web services whose functionalities fit with those required, it is necessary to then perform a semantic matching of the published services functional properties with those required and specified in the description of the abstract business process activities. Finally, to identify the most appropriate Web services which satisfy the enterprise constraints and preferences, this phase uses the non-functional properties specified in the modeling phase as criteria used to select services. Two algorithms for matching and selection are thereby necessary to perform Web services' dynamic discovery.

The matching algorithm aims to determine, for each activity of the required abstract process, a list of Web services providing the desired functionality. To optimize this operation, we propose to proceed in a hierarchical manner. The idea is to semantically compare each element specified in the functional model of the abstract activity with its correspondent extracted from the SAWSDL file, following a determined order to avoid inutile comparisons. This hierarchical order is shown in Figure 9 which illustrates the mapping of the elements described in the abstract activity's functional model with those of Web service, as well as the order in which the matching should be performed. The semantic concepts, corresponding to the business domain, functionality, inputs and outputs of the required abstract activity, should respectively be compared in this order with the SAWSDL file contents tagged *modelReference*, and corresponding to the interface, operation, inputs and outputs elements.

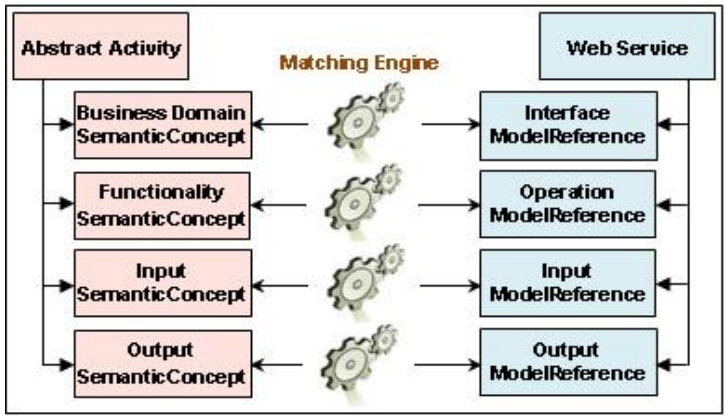

Fig. 9. Mapping between an abstract activity and Web service elements.

Besides, the semantic concepts do not usually perfectly match. There is often no exact matching between the matched concepts. To broaden the scope of the possible results by also covering the approximately similar concepts, we propose to enhance the matching algorithm by semantic relaxation mechanisms. This algorithm could, indeed, use the Edge-Counting Distance method (Hatzi et al., 2009) to calculate the semantic distance between two ontological concepts. This method calculates the distance between two concepts by counting the number of arcs of the shortest path connecting them in a given ontology. Calculating this metric allows determining the degree of similarity between two semantic concepts that can be either absolute or approximate.

Furthermore, data structures constituting the inputs/outputs types must be matched to determine the mapping between the data types used by an abstract process and those adopted by an invoked Web service. Calculating the semantic similarity, between the different concepts mentioned above, allows identifying Web services that match with the abstract business processes. However, the data types of their inputs and outputs can be heterogeneous. This can imply difficulties when integrating the various Web services that must interact to perform the different required business processes. To overcome this problem, it would also be necessary to make a matching of the inputs and outputs' data structures represented as XML schemas. We consider that a data type of a parameter used in an abstract business process and that corresponding to a parameter used by a published Web service are similar if they are equivalent or if the parameter's data type of the Web service covers that of the abstract process. When the matched data structures are heterogeneous, the matching algorithm should integrate data mediation mechanisms that make their mapping more flexible. This mapping would be useful for managing conflicts due to the heterogeneity of data exchanged during the Web services' invocation operation.

Finally, the selection algorithm adds a significant value to the discovery phase by specifying, among the list of discovered candidate services, those which suit the best to automate the enterprise's business processes. To perform the selection step, we propose to exploit the WS-Policy framework's abilities. This framework allows evaluating policies, and thereby, the non-functional properties of the discovered Web services, by comparing them with those related to the abstract business processes. The selection algorithm would, indeed, be important for ranking discovered Web services that are similar from the functional point of view. This ranking, based on the degree of their matching to the specific constraints and preferences of the enterprise, allows therefore refining the results of matching process.

## 6. Future research

The issue of modeling and integrating business processes using Web services is not limited to this work. There are several opportunities for future research that stem from the application of our approach. For instance, future works could build on the posited metamodel using Model Driven Architecture (MDA) to develop UML profiles for describing business processes. To explain the impact of this proposal, we recall that services' discovery is based on the matching of business processes properties with published Web services' elements. This imposes an issue due to the fact that these properties are encapsulated in UML diagrams that cannot be directly matched. To face this, we think that MDA techniques can be used in order to generate business processes' description files whose the syntactic structure is fully compatible, and thereby easily matched with that of Web services. In fact, similarly to our previous works in the Web services area (Belouadha et al., 2010a, 2010b; Omrana et al., 2010), we are actually considering to elaborate UML profiles in line with our metamodel and transformation rules in order to enable generating SAWSDL processes' files. This presents two advantages: on the one hand, the exploration of these files during the matching process would be easy, and on the other hand, its use for generating the BPMN diagram will enable generating a BPEL file which is patterned on Web services.

Second, future research can also be undertaken into the business ontologies to enable the semantic matching of business processes, and subsequently their dynamic discovery and integration. In fact, our approach uses SAWSDL principles in order to describe the business processes semantics. However, these principles are limited to semantic annotations. Indeed, the proposed metamodel only annotates the description elements used to design business processes and do not describe their semantics. For example, an ontology URL (Uniform Resource Locator) can be used to semantically annotate a given element describing a business process. Proposing sectorial ontologies or updating the existing ones in order to give semantic details concerning the diverse concepts related to business processes, particularly those considered in our metamodel, is important for supporting semantic processing such as the matching process. The required ontologies must consider the own characteristics of a given sector and give semantic relationships of different concepts, especially, business goals and functionalities, as well as the inputs and outputs of the related business processes. It can be built on the findings of the two main business schemas: North American Industry Classification System (NAICS) and Universal Standard Products and Services Classification (UNSPSC).

Lastly, future research can also build on our approach findings by shifting the emphasis more to the context-aware mobile business processes. The aim is to enable processes running in mobile environments benefiting from the potential of mobile devices. For instance, commercial agents often need to receive customers' orders on their cellular phones in order to activate the supply process and thereby rapidly satisfy the clients' requests. Thus, the main question remains: could contextual information, related to mobile devices properties (e.g., screen size and processor capacity) and users' preferences (e.g., preferred image or file formats), be considered at the business processes' design-time? In fact, our metamodel is based on Web services' W3C standards which are extensible. Thereby, we think that it will be possible to extend the proposed UML metamodel, and subsequently to obtain processes' description files including necessary contextual information. This information is considered as criteria for selecting appropriate mobile services which meet users' preferences and can run on specific mobile devices.

## 7. Discussion and conclusion

The Weakness of classical approaches for modeling and integrating business processes and the incessant need to create dynamic collaborative environments have consequently oriented the researchers in this area to explore the Web services. The emergence and the strengths of this technology have made it a prominent technology suitable for business process management and enterprise application integration. The interest of Web services to e-business has been discussed and confirmed by many researchers (Leymann et al. 2002; Albrecht et al., 2005; Zhang, 2005). This is what we have also addressed in this chapter.

Today, most enterprises use Web services to perform a part of their business processes, but they do not worry about how to design these processes, so that they become performed by Web services, and how to dynamically integrate them. However, this aspect is crucial to achieving complex business goals in a changing and evolving environment, and also being able to analyze and manage the enterprise business processes. Thereby, the approach we have proposed focuses on the use of Web services for process management in e-business. A number of recent works have been interested to this research orientation using different languages and formalisms. Papazoglou and Yang present a design methodology, based on WSDL and WSFL, for Web services and business processes (Papazoglou & Yang, 2002). They show how business process should be described so that services can be identified. Martens proposes a Petri nets-based method for modeling and analyzing business processes (Martens, 2005). Decker et al. propose an extension of BPEL to enable defining business process choreographies which are based on Web services (Decker et al., 2007). Gorton et al. present a workflow-based approach for business process modeling. This approach, founded on SOA, uses a simple graphical notation and the policy language Appel (Adaptable and Programmable Policy Environment and Language) to describe business processes, and aims to assemble and orchestrate available services in the business process (Gorton et al., 2009).

One of the advantages of our approach is that it enables to describe all business process aspects: functional, behavioral, but also semantic and non-functional. It is also completely aligned with the standards. To model and integrate business processes, it opts for the OMG standards, UML and BPMN, and the OASIS standard BPEL. Furthermore, it adopts design principles founded on the use of W3C standards, namely, WSDL, SAWSDL and WS-Policy. This leads to a business processes description which well fits with the standardized description of Web services. This compatibility of description allows, on the one hand, avoiding problems related to the heterogeneity of the description of the matched elements.

On the other hand, it allows leading to reliable discovery results since it could be possible to match all the concerned elements with their exact corresponding. Moreover, given the fact that the BPMN standard is directly mapped to BPEL, the use of this standard for orchestrating the activities of a business process or integrating many business processes allows to have business process files which are directly executable. It is also worth noting that the W3C standards used in our approach are extensible. This allows extending the proposed business process models in order to take into consideration unforeseen or specific aspects of business processes. Finally, our approach is a Web services-based and model-driven approach. It combines models and mechanisms of Web services' dynamic discovery to model and integrate business processes. The aim is not to show how to integrate a given Web service to legacy applications, when it is necessary to invoke it, but to define a way to develop enterprise information systems which can evolve and adapt to change, using agile business processes. Nevertheless, although our approach has advantages, it also imposes some limitations. First, it is only applied when business processes can be performed using Web services. Second, it can lead to a high cost in terms of dynamic discovery time due to the huge and evolving number of published Web services in Internet. Lastly, it assumes that the explored Web services are published in repositories according to new standards, namely, SAWSDL and WS-Policy.

## 8. References

Albrecht, C.C.; Dean, D.L. & Hansen, J.V. (2003). Market Place and Technology Standards for B2B, Ecommerce: Progress and Challenges. *MISQ Special Issue Workshop, MISQ Special Issue Workshop: Standard Making: A Critical Research Frontier for Information Systems, International Conference in Information Systems*, pp. 188-209, Seattle, Washington, USA, December 12-14, 2003

Albrecht, C.C.; Dean, D.L. & Hansen, J.V. (2005). Marketplace and Technology Standards for B2B, E-commerce: Progress, Challenges, and the State of the Art, *Information & Management*, Vol. 42, No. 6, (September 2005), pp. 865–875, ISSN 0378-7206

Belouadha, F.-Z. & Roudiès, O. (2008). Towards a UDDI based Semantic Solution for the E-business. *Proceedings of the 3rd IEEE International Conference on Information and Communication Technologies: from Theory to Applications ICTTA'08*, ISBN 978-1-4244-1752-0. Damascus, Syria, April 7-11, 2008.

Belouadha, F.-Z.; Omrana, H. & Roudiès, O. (2010a). A Model-driven Approach for Composing SAWSDL Semantic Web. *International Journal of Computer Science Issues (IJCSI)*, Vol. 7, Issue 2, No. 1, (March 2010), pp. 7-15, ISSN (Online) 1694-07 84, ISSN (Print) 1694-08 14, Retrieved from http://www.ijcsi.org/papers/IJCSI-Vol-7-Issue-2-No-1.pdf.

Belouadha, F.-Z.; Omrana, H. & Roudiès, O. (2010b). A MDA Approach for Defining WS-Policy Semantic Non-functional Properties. *International Journal of Engineering Science and Technology (IJEST)*, Vol. 2, Issue 6, (June 2010), pp. 2164-2171, ISSN 0975-5462, Retrieved from http://www.ijest.info/docs/IJEST10-02-06-115.pdf.

Brandenburg, H. & Wojtyna, J.-P. (2003). *L'approche Processus, Mode d'Emploi.* (2nd edition), Editions d'Organisation, ISBN 2-7081-3482-5, Paris, France

Charroux, B.; Osmani, A. & Thierry-Mieg, Y. (2009). *UML 2, Pratique de la Modélisation*, (3rd edition), Pearson Education France, ISBN 978-2-7440-7466-0, Paris, France

Chauvet, J.-M. (2002). *Services Web avec SOAP, WSDL, UDDI, ebXML*, Eyrolles, ISBN 978-2-212-11047-0, Paris, France

Chinnici, R.J.; Moreau, J.; Ryman, A. & Weerawarana, S. (June 2007). Web Services Description Language (WSDL) Version 2.0 Part 1: Core Language. W3C recommendation, In: *W3C Website*, 13.09.2011, Available from http://www.w3.org/TR/wsdl20/

Decker, G.; Kopp, O.; Leymann F. & Weske M. (2007). BPEL4Chor: Extending BPEL for Modeling Choreographies, Proceedings of IEEE International Conference on Web Services (ICWS 2007), pp. 296-303, 2007. ISBN 0-7695-2924-0, Salt Lake City, Utah, USA, July 09-13, 2007

Farrell, J. & Lausen, H. (August 2007). Semantic Annotations for WSDL and XML Schema, W3C recommendation, In: *W3C Website*, 13.09.2011, Available from http://www.w3.org/TR/sawsdl/

Gabay, J. & Gabay, D. (2008), *UML 2 Analyse et Conception : Mise en Oeuvre Guidée avec Etudes de Cas*, Dunod, ISBN 978-2-10-051830-2, Paris, France

Gorton, S.; Montangero, C.; Reiff-Marganiec, S. & Semini, L. (2009). StPowla: SOA, Policies and Workflows, In: *Service-Oriented Computing – ICSOC 2007 International Workshops, Vienna, Austria, September 17, 2007, Revised Selected Papers, LNCS Vol. 4907*, E. Di Nitto & M. Ripeanu (Ed(s).), pp. 351-362, Springer-Verlag, ISBN 978-3-540-93850-7, Berlin, Heidelberg, Germany, Retrieved from http://www.springer.com

Hammer, M. & Champy, J. (1993). *Le Reengineering – Réinventer l'Entreprise pour une Amélioration Spectaculaire de ses Performances*, Dunod, ISBN 2-10-002027-7, Paris, France

Hatzi, O.; Meditskos, G.; Vrakas, D.; Bassiliades, N.; Anagnostopoulos, D. & Vlahavas, I. (2009). Semantic Web Service Composition using Planning and Ontology Concept Relevance, *Proceedings of the IEEE/WIC/ACM International Joint Conference on Web Intelligence and Intelligent Agent Technologies (WI-IAT 2009)*, pp. 418-421, ISBN 978-0-7695-3801-3, Milan, Italy, September 15-18, 2009

Leymann, F.; Roller D. & Schmidt, M.T. (2002). Web Services and Business Process Management, *IBM Systems Journal*, Vol. 41, No. 2, pp. 198-211, ISSN 0018-8670

Lissandre, M. (1990). *Maîtriser SADT*. Editions Armand Colin, ISBN 2-200-42022-6, Paris, France

Lorino, P. (1991). *Le Contrôle de Gestion Stratégique, la Gestion par les Activités*, Dunod, ISBN 978-2-10-003080-4, Paris, France

Lorino, P. (1995). Le Déploiement de la Valeur par les Processus, *Revue Française de Gestion*, No. 104, (June-August 1995), pp. 55-71, ISSN 0338-4551

Martens, A. (2005). Analyzing Web Service based Business Processes. In: *Fundamental Approaches to software Engineering, 8th International Conference, FASE 2005, Held as Part of the Joint European Conferences on Theory and Practice of Software, ETAPS 2005, Edinburgh, UK, April 4-8, 2005 Proceedings, LNCS, Vol. 3442*, M. Cerioli (Ed.), pp. 19-33, Springer-Verlag, ISBN 978-3-540-25420-1, Berlin Heidelberg, Germany, Retrieved from http://www.springer.com

Morley, C. ; Bia-Figueiredo, M. & Gillette, Y. (2011). *Processus Métiers et Systèmes d'Information : Gouvernance, Management, Modélisation*, (3rd edition), Dunod, ISBN 978-2-10-055705-9, Paris, France

Mullenders, A. (2009). e-DRH: Outil de Gestion Innovant. la Théorie - les Progiciels - le Cadre Juridique. Groupe De Boeck s.a., ISBN 978-2-8041-5975-7, Bruxelles, Belgique

Octo Technlogy. (1999). Livre Blanc de l'EAI - Intégration des Applications d'Entreprise, In: *Scribd*, 13.09.2011, Available from http://www.scribd.com/doc/6964800/octocom-livre-blanc-eai

OASIS. (April 2007). Web Services Business Process Execution Language Version2.0, OASIS standard, In: *OASIS Website*, 13.09.2011, Available from http://docs.oasis-open.org/wsbpel/2.0/OS/wsbpel-v2.0-OS.html

OMG. (January 2011). Business Process Model and Notation (BPMN), Version 2.0, OMG Document, In: *OMG Website*, 13.09.2011, Available from
http://www.omg.org/spec/BPMN/2.0/PDF

Omrana, H.; Belouadha, F.-Z. & Roudiès, O. (2010). A MDA Approach for Describing Web Services Policies. *Proceedings of the 3rd IEEE International Conference on Web and Information Technologies (ICWIT'10)*, pp. 449-460, ISBN 978-9954-9083-0-3, Marrakech, Morocco, June 16-19, 2010.

Papazoglou, M. & Yang, J. (2002). Design Methodology for Web Services and Business Processes. In: *Technologies for E-Services, Third International Workshop, TES 2002, Hong Kong, China, August 23–24, 2002 Proceedings, LNCS, Vol. 2444*, G. Goos, J. Hartmanis & J. van Leeuwen (Ed(s).), pp. 54-64, Springer-Verlag, ISBN 3-540-44110-7, Berlin Heidelberg, Germany, Retrieved from http://www.springer.com

Smith, H.; Neal, D.; Ferrara, L. & Hayden, F. (2002). The Emergence of Business Process Management, Report by CSC's Research Services, Version 1.0, In: *Alarcos Website*, N. Morgan & A. Pappenheim (Ed(s)),13.09.2011, Available from
http://alarcos.inf-cr.uclm.es/doc/psgc/doc/lec/parte4b/csc-emergenceBPM.pdf

Tarondeau, J.-C. (1998). De Nouvelles Formes d'Organisation pour l'Entreprise : la Gestion par les Processus, *Les Cahiers Français*, No. 287, (July/September 1998), pp. 39-46. ISSN 00080217

Truman, G.E. (1998). An Empirical Appraisal of EDI Implementation Strategies. *International Journal of Electronic Commerce*, Vol. 2, No. 4, (June 1998), pp. 43-70, ISSN 1086-4415

Vedamuthu, A.; Orchard, D.; Hirsch, F.; Hondo, M.; Yendluri, P.; Boubez, T. & Yalçinalp, Ü. (September 2007). Web Services Policy 1.5 – Framework, W3C recommendation, In: *W3C Website*, 13.09.2011, Available from http://www.w3.org/TR/ws-policy/

WFMC, Work Group 1. (July 1998). Workflow Management Coalition, Interface 1: Process Definition Interchange Organisational Model, Document Number WfMC TC-1016-O, In: *WFMC Website*, 13.09.2011, Available from
http://www.wfmc.org/Download-document/WfMC-TC-1016-O-Interface-1-Process-Definition-Interchange-Organizational-Model.html

Zhang, D. (2005). Web Services Composition for Process Management in E-Business. *Journal of Computer Information Systems*. Vol. XLV, No. 2, (Winter 2005), pp. 83-91.